\documentclass[10pt]{article}
\usepackage{spconf,amsmath,amssymb,graphicx,booktabs,microtype,balance,tikz, enumitem}
\usetikzlibrary{arrows.meta,positioning}
\PassOptionsToPackage{hyphens}{url}
\usepackage[hidelinks]{hyperref}
\usepackage{CJKutf8}
\makeatletter
\let\kunqu@makecaption\@makecaption
\long\def\@makecaption#1#2{%
  \kunqu@makecaption{#1}{#2}%
  \def\kunqu@tabletype{table}%
  \ifx\@captype\kunqu@tabletype\vskip 6pt\relax\fi}
\makeatother
\newcommand{\zh}[1]{{\begin{CJK*}{UTF8}{gkai}#1\end{CJK*}}}

\title{KunSet: A Corpus and Benchmark for Lyrics Alignment \\ and Rhyme Retrieval in Kunqu Opera}

\name{\begin{tabular}{@{}c@{}}
Lejun Min\textsuperscript{2*}, Siqi Chen\textsuperscript{2*}, Delong Wang\textsuperscript{1*}, Tong Zheng\textsuperscript{1*}, Yunqi Lu\textsuperscript{2}, Zhishan Zhou\textsuperscript{1}, Ke Chen\textsuperscript{3}\\
Yuting Zhong\textsuperscript{1}, Junrong Li\textsuperscript{4}, Shengran Jin\textsuperscript{5}, Beixuan Pei\textsuperscript{6}, %%Chris Chafe\textsuperscript{2},
Jun Chen\textsuperscript{1\textdagger}, Yang Li\textsuperscript{1\textdagger}
\end{tabular}}
\address{
\textsuperscript{1}Peking University\quad
\textsuperscript{2}Stanford University\quad
\textsuperscript{3}University of California San Diego\\
\textsuperscript{4}Chinese Opera Channel CCTV\quad
\textsuperscript{5}Indiana University\quad
\textsuperscript{6}Sun Yat-sen University
\thanks{* These authors contributed equally to this work.}
\thanks{\textdagger\  Corresponding author.}
}

\begin{document}
% Use the template default 10 pt text and normal line spacing.
\maketitle
\ninept

\begin{abstract}
% Kunqu, among the oldest extant forms of Chinese opera, has been digitized mainly as recordings and texts, which lack the syllable-level alignment and phonological information needed for close analysis of performance. We present KunSet, a corpus of 281 Kunqu recordings (155.2 h; 106.3 h annotated) with 482,164 syllable-level timestamps linked to phonological attributes from Yunxue Lizhu, a Qing-dynasty rhyme book that we digitized into a searchable database. Timestamps were preannotated by an iteratively fine-tuned forced aligner and corrected by over 200 trained annotators under two-round review; the resulting annotations agree with two independent blind annotations as closely as the blind annotations agree with each other. We benchmark two tasks. For syllable-level lyrics alignment, fine-tuning on KunSet with a proposed acoustic anchor decoder halves the error of a pretrained aligner (MAE $391\to 195$ ms). For same-rhyme retrieval, phonological supervision on frozen HuBERT features raises P@10 from 11.4\% to 17.2\%. We release KunSet and model checkpoints.
Kunqu is one of the oldest extant forms of Chinese opera, yet its digitization remains scarce and difficult to annotate, with two challenges standing out: first, close analysis of Kunqu performance demands syllable-level lyrics alignment; second, analyzing Kunqu pronunciation requires its phonological attributes (e.g., rhyme) to be linked with each syllable. To address these challenges, we present KunSet, the first Kunqu performance corpus comprising 482,164 fine-grained annotations of syllable-level lyrics and phonological attributes in total media duration of 155.2 h with 106.3 h annotated. To ensure annotation quality, KunSet was preannotated by an iterative ML-assisted pipeline and corrected by over 200 trained annotators via a purpose-built online annotation system under a two-round review process. To assess KunSet as a music information retrieval (MIR) benchmark, we evaluate two tasks: lyrics alignment, by finetuning a forced-alignment model with an acoustic anchor decoder; and same-rhyme retrieval, using frozen Chinese HuBERT features. Models built with KunSet outperform their baselines. We release KunSet and model checkpoints to support MIR research on traditional-music corpora\footnote{\url{https://github.com/Kunyuan-Project/KunSet}}.  
% Studying Kunqu requires close listening to individual sung characters and comparing their sounds across performances. We present KunyuanSet, comprising the first character-aligned Kunqu performance corpus and the first Kunqu character-rhyme dataset. This performance corpus contains 283 recordings of length 156.5h, with 107.3h covered by 486,853 annotated characters from the Kunqu Opera Art Canon. To support large-scale Kunqu character alignment, we contribute a forced alignment model finetuned on our corpus with a novel acoustic anchor decoder. We also developed an online annotation system supporting over 200 participants in correcting forced-alignment preannotations and completing two rounds of review. We structure the historical pronunciation reference \textit{Yunxue Lizhu} into a character-rhyme database of 10,235 entries with searchable phonological attributes. For same-rhyme audio retrieval, traditional rhyme supervision over frozen Chinese HuBERT features achieves higher rhyme retrieval precision compared to modern Mandarin-final supervision, demonstrating the value of traditional phonological knowledge for Kunqu audio retrieval.
\end{abstract}


\begin{figure*}[t]
\centering
\includegraphics[
    width=\textwidth,
    trim={1cm 0cm 1cm 0cm},
    clip
  ]{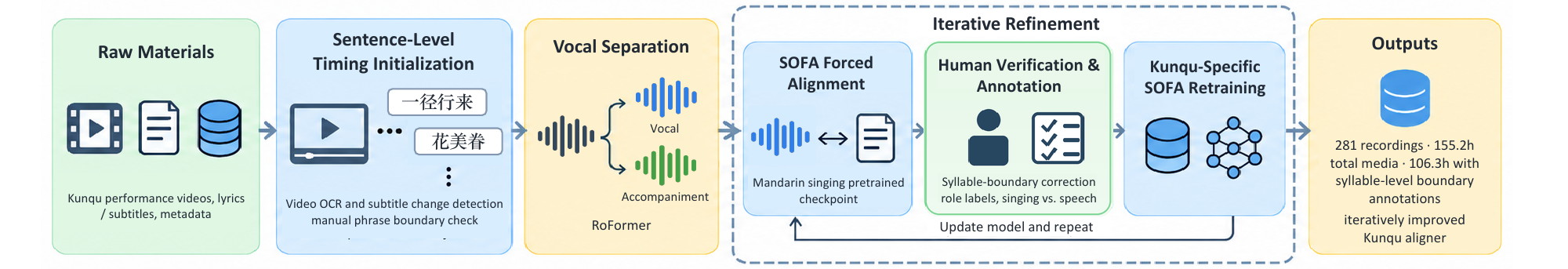}
\caption{Annotation pipeline for the syllable-aligned corpus.}% Subtitle OCR initializes sentence timing, Mel-Band RoFormer separates vocals, and SOFA produces syllable preannotations. Annotators correct the boundaries and add delivery-mode and role labels, and the corrected recordings are used to finetune the aligner for the next preannotation round. Section~\ref{sec:alignment}).}
\label{fig:pipeline}
\end{figure*}

\section{Introduction}
\label{sec:intro}

Honored as the ``teacher of a hundred Chinese theatrical genres'', Kunqu was proclaimed a Masterpiece of the Oral and Intangible Heritage of Humanity by UNESCO in 2001~\cite{unescoKunqu}. Over six centuries of development it accumulated a vast repertoire of plays and techniques transmitted almost entirely by oral instruction and imitation, many of which have already been lost~\cite{unescoKunquTraining}. Its digitization and close analysis are therefore urgent. A central unit for such analysis is the syllable: Kunqu melodies are set to follow the tones of the lyrics, and each syllable is articulated through its head, belly and tail~\cite{shenDuqu}. Studying these practices across performances requires both syllable-level timing and phonological information.

One of the most important digitization efforts, \textit{Kunqu Opera Art Canon}~\cite{artcanon}, preserves texts, scores and audiovisual records of 297 expert-selected performed excerpts. KunquDB~\cite{kunqudb} further provides sentence-level timestamps and speaker metadata, but lacks finer syllable boundaries. Another branch of Kunqu preservation focuses on its phonological system, which informs how syllables are sung and articulated~\cite{shenDuqu}. Shen Chenglin’s \textit{Yunxue Lizhu} (\zh{韵学骊珠}), compiled in 1792, provides a systematic Kunqu phonological taxonomy of Chinese characters~\cite{shenYunxue,shi2014}. An existing digital interface supports source-page lookup~\cite{gongchepuYxlz}, yet phonological attributes are not directly searchable. \textit{Nine Modes Manual Online}~\cite{schoenberger2021ninemodes} annotates tone and yin-yang for a historical Kunqu melody manual, but does not link them to performance recordings. Together, these resources leave two gaps: 1) syllable-level alignment of sung and spoken lyrics to performance recordings, and 2) phonological attributes linked to each syllable. Addressing these requires large-scale annotation and structured phonological digitization.

Recent developments in forced alignment provide a basis for obtaining syllable boundaries at scale. MFA~\cite{mfa} addresses speech alignment, while CTC-based methods~\cite{vaglio2020} and SOFA~\cite{sofa} target singing voices. However, forced alignment models pretrained on speech or modern singing perform poorly on Kunqu, where sustained melismatic syllables and stylized pronunciation lie outside their training distribution.

In this work, we close both gaps with three contributions:
\begin{itemize}[leftmargin=*]
    \item We present KunSet, to our knowledge, the first syllable-aligned Kunqu performance corpus with linked phonological attributes, comprising 482,164 syllable-level timestamps over 106.3 h of sung and spoken lyrics in 281 recordings, supporting comparison between actual interpretation and traditional phonological knowledge.
    
    \item We design a curation pipeline: an iteratively fine-tuned SOFA model preannotates syllable boundaries, with corrections by over 200 trained annotators under two-round review in a purpose-built system, validated through two independent blind annotations and acoustic analyses; and a searchable digitization of \textit{Yunxue Lizhu} that links syllables to candidate phonological attributes.
    
    \item We build two Kunqu-specific MIR systems and benchmark them on KunSet: a fine-tuned SOFA model with a proposed acoustic anchor decoder for syllable-level lyrics alignment, and a same-rhyme retrieval model trained with phonological supervision on frozen HuBERT features. Both outperform their baselines, demonstrating KunSet's value for Kunqu MIR research.
\end{itemize}

% \noindent We close both gaps with three contributions:
% \begin{itemize}[leftmargin=*]
%     \item KunSet, to our knowledge the first syllable-aligned Kunqu performance corpus with linked phonological attributes, comprising 482,164 syllable-level timestamps over 106.3 h of sung and spoken lyrics in 281 recordings, supporting close analysis and comparison between actual interpretation and traditional phonological knowledge.
%     \item A curation pipeline addressing both gaps: an iteratively fine-tuned SOFA~\cite{sofa} model preannotates syllable boundaries, which over 200 trained annotators correct under two-round review in a purpose-built system, with consistency supported by two independent blind annotations and acoustic analyses; and a searchable digitization of Yunxue Lizhu (10,235 entries) that links syllables to candidate phonological attributes.
%     \item A benchmark with two tasks: syllable-level lyrics alignment, measuring how well KunSet adapts aligners to Kunqu, where fine-tuning with a proposed acoustic anchor decoder halves the error of a pretrained aligner; and same-rhyme retrieval, testing whether the linked phonological categories are recoverable from performance audio, where traditional phonological supervision of frozen HuBERT features raises P@10 by 5.8 points.
% \end{itemize} 

\section{Data Collection and Annotation}
\label{sec:corpus}
In this section, we will first introduce the data format, label ontology, and statistics of KunSet. Then we will introduce how we design the annotation and review pipeline to leverage both ML models and human correction.

\subsection{Dataset Format and Label Ontology}
\label{sec:dataset-format}
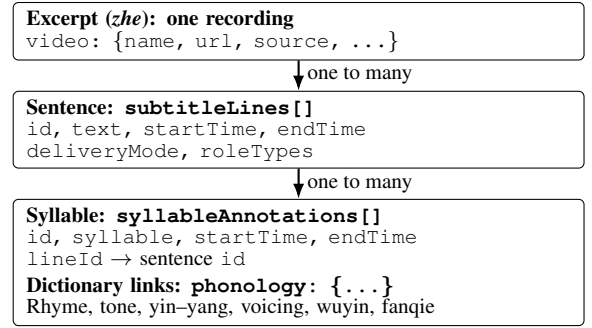
\begin{figure}[t]
\centering
% Logical hierarchy; field names match ProjectData and reviewed JSON exports.
% The sentence and character arrays are siblings in the serialized JSON.
\begin{tikzpicture}[
  font=\fontsize{8}{8.5}\selectfont,
  levelbox/.style={draw,rounded corners=2pt,align=left,
    text width=7.2cm,inner xsep=5pt,inner ysep=3pt},
  link/.style={-{Latex[length=2mm]},thick}]
\node[levelbox] (excerpt) {
  \textbf{Excerpt (\textit{zhe}): one recording}\\
  %Annotation JSON\\
  \texttt{video: \{name, url, source, \ldots\}}};
\node[levelbox,below=0.4cm of excerpt] (sentence) {
  \textbf{Sentence: \texttt{subtitleLines[]}}\\
  \texttt{id, text, startTime, endTime}\\
  \texttt{deliveryMode, roleTypes}};
\node[levelbox,below=0.4cm of sentence] (syllable) {
  \textbf{Syllable: \texttt{syllableAnnotations[]}}\\
  \texttt{id, syllable, startTime, endTime}\\
  \texttt{lineId} $\rightarrow$ sentence \texttt{id}\\[2pt]
  \textbf{Dictionary links: \texttt{phonology: \{\dots\}}}\\
  %\texttt{candidateEntryIds[], selectedEntryId}\\
  %\texttt{attributes, attributeStatus}\\
  Rhyme, tone, yin--yang, voicing, wuyin, fanqie};
\draw[link] (excerpt.south) -- node[right]{one to many} (sentence.north);
\draw[link] (sentence.south) -- node[right]{one to many} (syllable.north);

\end{tikzpicture}
\caption{Excerpt--sentence--syllable annotation hierarchy (selected fields). Sentence and syllable annotations are linked by \texttt{lineId}, with all times referring to the recording timeline. Each syllable links to dictionary candidates through \texttt{phonology}.}
\label{fig:hierarchy}
\end{figure}

KunSet is derived from 281 high-quality Kunqu excerpts in \textit{Kunqu Opera Art Canon}. The label ontology of KunSet is illustrated in Fig.~\ref{fig:hierarchy} as excerpt--sentence--syllable hierarchy. All labels are represented in JSON format:

\begin{itemize}[leftmargin=*]
    \item Excerpt: the video recording of one performed Kunqu excerpt (\textit{zhe}), with a 720$\times$576, 25\,fps video track and stereo audio at 44.1\,kHz. 
    \item Lyric sentence: one sentence of sung lyrics or spoken dialog in the recording, with its start time, end time, the stage character(s) delivering it, its delivery mode (\textit{sung} or \textit{spoken}), and the transcript.
    \item Syllable: the start time, end time, and corresponding phonological attributes of every Chinese character (one syllable each) within a sentence.
\end{itemize}

In Kunqu opera, each Chinese character (\zh{字}) is one syllable and thus an unambiguous unit of lyrics alignment. We associate each lyric sentence with its constituent syllables through a unique \texttt{lineId}, preserving sentence-level text, delivery mode, and role information. Both sentence and syllable annotations are defined by onset and offset times on the recording timeline. Each syllable annotation additionally includes a \texttt{phonology} field linking it to candidate readings in \textit{Yunxue Lizhu}. We retain all dictionary candidates and populate each attribute only when there is a single candidate or all candidates agree on a non-missing value; unresolved attributes remain unset.

\begin{figure}[t]
\centering
\includegraphics[width=\columnwidth]{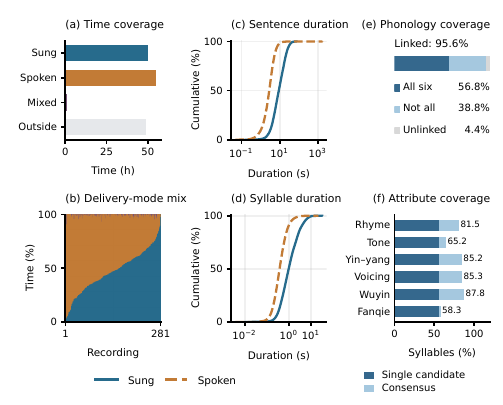}
\caption{Overview of 281 recordings. (a) Time by type; mixed denotes conflicting labels, outside denotes unannotated time. (b) Per-recording time composition, sorted by singing share. (c, d) Sentence and syllable duration CDFs (log-scaled). (e) All six attributes resolved, linked but not all resolved, or unlinked. (f) Attributes resolved by a single candidate or candidate consensus.}
\label{fig:corpus-profile}
\end{figure}

Summarized in Table~\ref{tab:corpus-stats} and Fig.~\ref{fig:corpus-profile}, the dataset consists of 281 recordings, with labeled singing and stage speech data. The annotated corpus is approximately balanced between singing (47.3\%) and stage speech (51.9\%), while individual recordings vary substantially in relative proportions. Sung sentences have a longer median duration than spoken
sentences (8.81\,s versus 2.77\,s), consistent with the sustained vocalization characteristic of Kunqu singing. Of the 482,164 syllable annotations, 461,164 (95.6\%) have dictionary candidates, and 273,996 (56.8\%) have unambiguous values for all six phonological attributes.

% Generated by statistics/corpus_statistics.py; do not edit numbers.
\begin{table}[t]
\centering
\caption{Descriptive statistics of the 281-recording KunSet corpus. Time coverage uses interval unions.}
\label{tab:corpus-stats}
\setlength{\tabcolsep}{3pt}
\begin{tabular}{lr}
\toprule
Measure & Value \\
\midrule
Recordings (the number of) & 281 \\
Media / lyrics-covered time (h) & 155.19 / 106.34 \\
Sung / spoken / mixed (h) & 50.30 / 55.15 / 0.90 \\
Sentence / syllable records & 78,505 / 482,164 \\
Distinct transcript characters & 4,357 \\
Sentence time coverage (known media) & 68.5\% \\
Phonology-linked / resolved syllable records & 461,164 / 273,996 \\
\bottomrule
\end{tabular}
\end{table}

\subsection{Annotation and Review Pipeline}

% paragraph 1: annotation strategy / motivation -- why the workflow and platform

To overcome the cost and difficulty of manual lyrics alignment, we adopt a preannotation-and-review workflow: initial syllable-level timestamps are generated automatically through an ML-assisted pipeline, corrected by one annotator, and then passed through two rounds of review by different reviewers.

% paragraph 2: preannotation pipeline
The preannotation pipeline is summarized in Fig.~\ref{fig:pipeline}. We first used VideOCR~\cite{videocr} to extract initial sentence timestamps and transcripts from the recordings, and Mel-Band RoFormer~\cite{wang2023melband} to separate vocals from accompaniment. We initialized the iterative process by manually annotating one Kunqu excerpt and fine-tuning SOFA's pretrained Mandarin singing checkpoint~\cite{sofa} on it. The adapted model then produced additional annotated excerpts, which were corrected and reviewed before being added to the fine-tuning set. Repeating this cycle expanded the training set from one to seven excerpts, and the resulting model generated initial timestamps for the remaining 274 recordings. This preannotation model is included in the alignment benchmark in Section~\ref{sec:alignment}.

% paragraph 3: annotation granularity
% We use syllable boundaries as the smallest annotated unit rather than phone boundaries for two reasons: (1) standard Pinyin does not accurately represent the phonetics of Kunqu stage pronunciation; (2) given this mismatch, consistently annotating phone boundaries would require specialized phonological expertise and would be impractical across a large annotator pool. As described in Section~\ref{sec:dataset-format}, each annotated Chinese character corresponds to one syllable in KunSet.

% paragraph 4: human correction, review, collaborative platform (include demo here)

An online annotation system is developed for over 200 trained annotators (demo available\footnote{\url{https://tinyurl.com/kunset-icassp-demo}}). The system displays video, sentence transcripts, and editable syllable intervals on a shared, zoomable timeline. Optional waveform and spectrogram displays support acoustic inspection, as in established tools such as Praat~\cite{boersma2001praat}. Delivery mode (singing or stage speech) and role labels are annotated separately at the sentence level. Project assignments, file-level permissions, and completion states are used to coordinate annotation and review across the annotator pool. Reviewers confirm annotations, leave time-linked comments, correct boundaries when necessary, and flag unresolved cases for further review. After correction and review, the annotations are exported in the JSON format described in Section~\ref{sec:dataset-format}.

\subsection{Searchable Phonological Database}
To link every syllable with traditional phonological attributes, we structured \textit{Yunxue Lizhu} into a searchable database that provides fanqie, tone, yin--yang, rhyme, wuyin, and voicing for each candidate entry. We commissioned Classical Chinese experts to transcribe the original print, and converted the transcription to JSONL and SQLite. Exact lookup returns all candidates, with recorded variants providing a fallback. The database contains 10,235 entries for 8,300 direct-lookup Chinese characters, including 1,664 characters with multiple entries and 2,520 variant-alias relations.

\subsection{Annotation Quality Assessment}
\label{sec:quality}

To assess annotation consistency and verify that the ML preannotation stage does not introduce bias, we compare the reviewed production annotations (M) on the 10 randomly selected test recordings with two independently produced blind annotations, A and B. For each blind annotation, a seeded monotonic time warp shifts sentence starts and ends by 1--3 seconds, using different seeds for A and B while preserving sentence order and shared boundaries and preventing adjacent-sentence overlap. Syllable intervals are then initialized by uniformly dividing each perturbed sentence in transcript order. The two versions are then independently corrected with two rounds of review.

\begin{table}[t]
\centering
\caption{Production (M) versus blind annotations (A/B), grouped by A's delivery labels. Onset and offset MAEs (ms) are recording-macro averages; lower is better.}
\label{tab:production-blind}
\fontsize{9}{10.5}\selectfont
\setlength{\tabcolsep}{2pt}
\begin{tabular}{@{}lrrrrrr@{}}
\toprule
 & \multicolumn{2}{c}{Overall $\downarrow$} & \multicolumn{2}{c}{Sung $\downarrow$} & \multicolumn{2}{c}{Spoken $\downarrow$} \\
\cmidrule(lr){2-3}\cmidrule(lr){4-5}\cmidrule(lr){6-7}
Pair & Onset & Offset & Onset & Offset & Onset & Offset \\
\midrule
M--A & 158.8 & 206.4 & 272.1 & 354.1 & 115.8 & 156.7 \\
M--B & 172.5 & 225.5 & 309.1 & 408.5 & 126.2 & 165.0 \\
A--B & 158.1 & 207.3 & 282.8 & 384.0 & 108.2 & 147.8 \\
\bottomrule
\end{tabular}
\end{table}

Using A as the text reference, we identify 15,710 syllable instances shared across all three versions. We compute onset and offset MAE on these matched instances, with recordings weighted equally (Table~\ref{tab:production-blind}). Overall, the M--A and M--B discrepancies are close to those between the two blind annotations, A--B, with MAEs differing by at most 18.14\,ms. This suggests that the reviewed production annotations achieve temporal consistency comparable to independently produced blind annotations. Across all three pairs, disagreement is greater for singing than speech and for offsets than onsets, suggesting that these patterns reflect shared annotation difficulty rather than a discrepancy specific to ML-assisted production.

\subsection{Acoustic Consistency Verification}

\begin{figure}[t]
\centering
\includegraphics[width=\columnwidth]{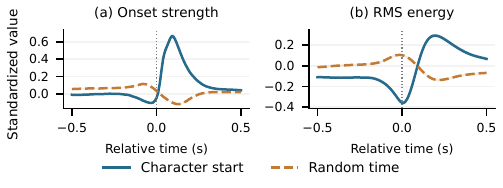}
\caption{Acoustics around syllable starts and random times in the same
sentence. Zero marks the sampled time; shading shows pointwise 95\%
confidence intervals from 2,000 recording-level bootstrap resamples.}
\label{fig:character-acoustics}
\end{figure}

We further evaluate annotation consistency by testing whether annotated syllable starts correspond to systematic acoustic changes. Across 281 recordings, we analyze syllable starts using source-separated vocals. Each start is compared with five other random times in the same sentence, at least 100\,ms from any annotated start. We compute onset strength (positive log-mel spectral flux) and RMS energy with librosa~\cite{librosa2015}, standardized within recordings and averaged with equal recording weights. Figure~\ref{fig:character-acoustics} shows that syllable starts exhibit sharp increases in onset strength and RMS energy from near-onset troughs, whereas random-time controls decline over the same interval. The 0--200\,ms onset-strength contrast is 0.47 standard deviations (recording-level bootstrap 95\% CI: 0.46--0.49), and remains 0.43 (0.42--0.45) after excluding sentence-initial syllables. This supports a corpus-level association between the annotated starts and acoustic transitions, including within sentences.

\section{Experiments}
To demonstrate how KunSet supports MIR research on traditional-music corpora, we benchmark two tasks: syllable-level lyrics alignment, to assess its value for adapting alignment models to Kunqu, and same-rhyme retrieval, to examine the benefit of traditional phonological supervision.

\subsection{Syllable-level lyrics alignment}
\label{sec:alignment}
\textbf{Data and models.} We evaluate syllable-level lyrics alignment on a recording-level split of 251 Train, 20 Dev and 10 Test recordings.
% (88.57h, 6.84h, 3.47h covered by lyrics, respectively)
The 10 Test recordings use blind annotation A as the reference. Models receive source-separated vocals by Mel-Band RoFormer~\cite{wang2023melband} and the corresponding sentence transcripts. We compare pretrained Mandarin-singing SOFA~\cite{sofa} with Kunqu-adapted SOFA fine-tuned on 7 excerpts (from the iterative preannotation stage) and 251 excerpts (full scale), each decoded using either the original decoder or the proposed acoustic anchor decoder. \\ 

\noindent \textbf{Training settings}\; For the 251-recording model, we fine-tune the acoustic backbone and prediction head for 9,000 steps with three seeds. We use AdamW optimizer with weight decay 0.1 and a one-cycle learning-rate schedule with peak rates of $6\times10^{-5}$ for the backbone and $3\times10^{-4}$ for the head. We use pretrained SOFA to supply within-syllable phone-boundary pseudo-labels, with a boundary-loss weight of 0.1 for training. For each seed and decoder, checkpoints are selected by recording-macro Dev boundary MAE.\\ 

\noindent \textbf{Acoustic anchor decoding.} Sustained syllables can have unclear edges but strong phone evidence within them. We design the decoder to select boundaries jointly with the strongest evidence frame in each candidate phone interval. Building on SOFA's phone-peak scoring~\cite{sofa}, it maximizes:
\begin{align}
\resizebox{0.92\columnwidth}{!}{$
F(\pi)={}\sum_{t=1}^{T}e_{t,\pi_t}+B(\pi)-M(\pi)\log 3
+\frac{2T}{S}\sum_{s:p_s\ne\mathrm{SP}}
\max_{t\in I_s(\pi)}e_{ts}$}
\label{eq:acoustic-anchor}
\end{align}
Here $\pi$ assigns $T$ frames to $S$ ordered transcript positions with phone labels $p_s$ and intervals $I_s(\pi)$; only silence (SP) may be skipped. $e_{ts}$ is frame-level phone evidence, $B(\pi)$ scores boundaries, and $M(\pi)\log 3$ penalizes entering positions. The peak term, weighted by $2T/S$, rewards intervals containing strong evidence for the expected phone. An exact $O(TS)$ search jointly selects anchors and boundaries, after which phone intervals are grouped into syllables. The decoder changes inference without changing acoustic training.\\ 

\noindent \textbf{Evaluation and results.}
We evaluate 11,735 internal syllable boundaries using MAE in milliseconds and Acc@200, the percentage within 200\,ms, both averaged equally across recordings. Results for the 251-excerpt adapted models are averaged over three seeds. Table~\ref{tab:main} shows complementary gains from corpus-scale adaptation and acoustic-anchor decoding. Fine-tuning on 251 excerpts substantially reduces MAE under both decoders, while acoustic anchors further reduce adapted-model MAE by 13.04\% and increase Acc@200 from 79.52\% to 81.69\%. The 251-excerpt model also clearly outperforms the seven-excerpt preannotation model under both decoders, showing continued benefit from the expanded reviewed corpus.

Although Table~\ref{tab:production-blind} uses a different boundary set, the comparison provides useful context: on sung sentences the adapted model's error (462.48\,ms) remains above human disagreement (272--409\,ms), whereas on spoken sentences it is of the same magnitude (123.77\,ms versus 108--165\,ms). This suggests that the main remaining alignment difficulty lies in Kunqu singing instead of stage speech.

% Generated by docs/validation/alignment-scale-mode-20260923/summarize.py.
\begin{table}[t]
\centering
\caption{Test alignment overall and by delivery mode. MAE (ms) and Acc@200 (\%) are recording-macro averages. Ours (FT251): 3-seed means, overall MAE seed SDs $\leq$4.5\,ms; Ours (FT7): one model.}
\label{tab:main}
\fontsize{7}{8}\selectfont
\setlength{\tabcolsep}{1pt}
\begin{tabular}{@{}llcccccc@{}}
\toprule
 & & \multicolumn{2}{c}{Overall} & \multicolumn{2}{c}{Sung} & \multicolumn{2}{c}{Spoken} \\
\cmidrule(lr){3-4}\cmidrule(lr){5-6}\cmidrule(lr){7-8}
Model & Decoder & MAE $\downarrow$ & Acc@200 $\uparrow$ & MAE $\downarrow$ & Acc@200 $\uparrow$ & MAE $\downarrow$ & Acc@200 $\uparrow$ \\
\midrule
Pretrained & Original & 391.30 & 61.93 & 797.83 & 48.78 & 288.63 & 64.72 \\
Pretrained & Anchor & 337.70 & 65.12 & 702.34 & 50.76 & 241.07 & 68.45 \\
Ours (FT7) & Original & 372.91 & 64.49 & 753.40 & 53.66 & 267.69 & 67.25 \\
Ours (FT7) & Anchor & 312.90 & 67.60 & 620.62 & 56.64 & 233.98 & 70.35 \\
Ours (FT251) & Original & 224.04 & 79.52 & 529.21 & 65.35 & 144.95 & 83.46 \\
Ours (FT251) & Anchor & \textbf{194.83} & \textbf{81.69} & \textbf{462.48} & \textbf{67.50} & \textbf{123.77} & \textbf{85.84} \\
\bottomrule
\end{tabular}
\end{table}

\subsection{Same-rhyme retrieval}
\label{sec:lexicon}
\textbf{Data and models.} Using the same 251/20/10 recording split and known-mode (singing/spoken) sentence selection as alignment, we retain instances with unambiguous traditional rhyme and modern Mandarin-final labels: 256,333 Train and 20,674 Dev instances, and 10,287 Test instances. All 29 rhyme groups have supported Test queries. A frozen Chinese HuBERT-base encoder~\cite{hsu2021hubert,chineseHubert} processes each clip at 16\,kHz. The mean and standard deviation of its final-layer frame representations form a 1,536-dimensional feature, standardized using Train statistics. A projection head maps $1536\!\rightarrow\!256\!\rightarrow\!128$ dimensions with ReLU, dropout 0.1 and output $L_2$ normalization. We compare traditional rhyme labels, modern Mandarin finals and traditional labels shuffled within each training recording; frozen HuBERT features without a projection head serve as the baseline.\\ 

\noindent \textbf{Training settings.} We train the projection head on each learned condition using supervised contrastive loss~\cite{khosla2020supcon} with temperature 0.1 and uniformly weighted positive pairs from different Chinese characters and recordings. Batches sample up to eight classes and 16 instances per class. We use AdamW with learning rate $10^{-3}$ and weight decay $10^{-4}$. All conditions train for 200 epochs of 400 steps with three seeds. We use Dev P@10 to select checkpoints before Test evaluation.\\ 

\noindent \textbf{Evaluation and results.}
Queries search the Train library by cosine similarity, excluding candidates with the same Chinese character identity or recording as the query. The primary metric, P@10, is the fraction of same-rhyme instances among the ten nearest neighbors, averaged equally across rhyme groups. Table~\ref{tab:retrieval} reports Test means and standard deviations over three seeds on the common query set. 

Traditional supervision achieves the highest P@10 at 17.16\%, exceeding modern Mandarin finals by 0.81 percentage points (95\% paired recording-bootstrap interval: [0.47, 1.18]) and frozen HuBERT by 5.80 percentage points; shuffled supervision performs substantially worse. The small performance gap between traditional rhyme supervision and modern Mandarin finals is consistent with substantial overlap in the phonological information captured by the two systems, while the larger gain over frozen features shows the benefit of phonological supervision for organizing performance audio by rhyme. Because retrieval is evaluated using traditional rhyme labels, the traditional--modern difference should be interpreted in light of this taxonomy match.

% Generated by ../scripts/render_hubert_test.py from frozen Test evidence.
\begin{table}[t]
\centering
\caption{Same-rhyme retrieval on the 10-recording Test pool: macro P@10 (\%), mean $\pm$ sample SD over three seeds.}
\label{tab:retrieval}
\setlength{\tabcolsep}{4pt}
\begin{tabular}{lr}
\toprule
Supervision & P@10 $\uparrow$ \\
\midrule
None (HuBERT) & 11.36 \\
Shuffled rhyme groups & 8.81\,\textpm\,0.07 \\
Modern Mandarin finals & 16.35\,\textpm\,0.14 \\
Traditional rhyme groups (ours) & \textbf{17.16}\,\textpm\,0.11 \\
\bottomrule
\end{tabular}
\end{table}

%\textbf{Analysis.}
%The P@10 advantage supports the usefulness of traditional rhyme supervision for retrieving denser sets of same-rhyme examples. Shuffled-label training falls below the frozen baseline, indicating that fitting a projection alone does not account for the improvement.

\section{Conclusion}
\label{sec:limits}
We presented KunSet, a Kunqu corpus comprising 281 recordings with 482,164 syllable-level annotations and links to phonological attributes from \textit{Yunxue Lizhu}. Experiments demonstrate improved lyrics alignment and the benefit of traditional phonological supervision for same-rhyme retrieval. The dataset, code, and model weights are released for research-purpose only, providing a foundation for MIR research on Kunqu.

% Future work will incorporate video understanding and traditional \textit{gongche} notation, linking performance visuals and musical scores to the aligned text and audio. Our goal is a hierarchical, cross-modal Kunqu database that connects characters, phrases and performed excerpts across these modalities, supporting research on pronunciation, melody and stage performance.

\clearpage

\section{Acknowledgements}
This study is funded by the School of Arts, Peking University, as part of the Kunyuan project.%We would like to thank Prof. Jaros{\l}aw Kapu\'{s}ci\'{n}ski and Prof. Rongfeng Li for their kind support.

% Four pages of self-contained technical content; references start on page five.

\balance
\begingroup
\emergencystretch=2em
\hyphenation{rec-og-ni-tion}
\bibliographystyle{IEEEbib}
\bibliography{refs}
\endgroup
\end{document}